\documentstyle[draft,epsf]{mn}

\def\hi{{\rm H\,{\sc i} }}
\def\hii{{\rm H\,{\sc ii} }}

\title{The Stellar Populations of Low Surface Brightness Galaxies}

\author[Bell et al.  ]
{
Eric F. Bell$^1$, Richard G. Bower$^1$, 
Roelof S. de Jong$^{1,3}$\thanks{Hubble Fellow}, 
Mark Hereld$^2$, and  \vspace{0.1cm} \\
{\LARGE Bernard J. Rauscher$^1$}\\
$^1$ Department of Physics, University of Durham, Science Labs, South Road, 
Durham DH1 3LE, UK \\
$^2$ Department of Astronomy and Astrophysics, 
University of Chicago, 5640 S. Ellis
Ave., Chicago, IL 60637, USA\\
$^3$ Steward Observatory, University of Arizona, 
949 N. Cherry Ave., Tucson, Arizona, 85719, USA}

\begin{document}
\date{Accepted by MNRAS Letters: \today}

\maketitle

\begin{abstract}
Near-infrared (NIR) $K'$ images of a sample of five low surface
brightness disc galaxies (LSBGs) were combined with optical
data, with the aim of constraining their star formation histories.
Both red and blue LSBGs were imaged to enable comparison of their
stellar populations.
For both types of galaxy strong colour gradients were 
found, consistent with mean
stellar age gradients.  Very low stellar metallicities were ruled out
on the basis of metallicity-sensitive optical-NIR colours.  
These five galaxies suggest that
red and
blue LSBGs have very different star formation histories and represent two
independent routes to low $B$ band surface brightness.  Blue LSBGs are
well described by models with low, roughly constant star formation
rates, whereas red LSBGs are better described by a `faded disc' scenario.  
\end{abstract}


\section{Introduction}

Low surface brightness disc galaxies (LSBGs; galaxies with $B$ band
central surface brightnesses fainter than 22.5 mag\,arcsec$^{-2}$) may
comprise up to $\sim\,1/2$ of the local galaxy population
\cite{mcgaugh1995a}.  Yet, due to their faintness compared to the
night sky, they have remained largely overlooked in most local galaxy
surveys.  Recently, the compilation of galaxy catalogues specially
designed to detect LSBGs (Schombert \& Bothun 1988; Impey et al.\ 1996;
O'Neil, Bothun \& Cornell 1997a) has
%\cite{schombert1988,impey1996,oneil1997} has
allowed systematic study of their properties.
Despite this, many aspects of their formation and evolution remain
poorly understood.  In particular, there is considerable
uncertainty regarding their star formation history (SFH).

The most widely studied LSBGs are blue
\cite{mcgaugh1994a,deblok1995}, indicating a young mean stellar age
and/or low metallicity (note that this may be a selection effect, as
most LSBG samples are selected from blue photographic survey plates).  
Their measured \hii region metallicities are
low at around or below $1/3$ solar abundance
\cite{mcgaugh1994b,ronnback1995,deblok1997ab}.
Morphologically, the best studied LSBGs appear to have discs, but
little spiral structure \cite{mcgaugh1995b}.  The
massive star formation rates (SFRs) in LSBGs are 
an order of magnitude lower than those of high surface
brightness (HSB) galaxies \cite{vdH93}.  \hi observations show that
LSBGs have large gas mass fractions, sometimes even approaching unity
\cite{deblok1996,mcgaugh1997}.  As yet, there have been no CO
detections of LSBGs, only upper limits on the CO abundances which
indicate that LSBGs have CO/\hi ratios significantly lower than those
of HSB galaxies \cite{schombert1990,deblok1997co}.  All of the above
strongly suggests that blue LSBGs are relatively unevolved, low mass surface
density, low metallicity systems, with roughly constant or even
increasing SFRs \cite{deblok1997sfr}.

However, recent work has demonstrated the existence of a red population of
LSBGs, with both central and outer optical colours compatible with
those seen in old stellar populations \cite{oneil1997,oneil1997b}.  
In this letter, we explore the differences between a sample of five
LSBGs (three blue and two red) taken from a larger ongoing study  
(note that we do not consider giant LSBGs, such as Malin 1, in this study).
Because of the age/metallicity
degeneracy inherent in optical broad-band colours, 
it is impossible to tell from the optical data alone 
exactly how the red LSBGs differ
from their blue counterparts, and what drives their optical colour
gradients \cite{deblok1995}.  NIR images, in conjunction with optical
data, offer the first chance to break this degeneracy, and constrain plausible
SFHs.

\section{Observations and data reduction}

Our sample is taken from de Blok et al.\ \shortcite{deblok1996} (F561-1,
F563-V2 and F568-3)
and O'Neil et al.\ \shortcite{oneil1997} (C3-2 and N10-2) 
and is selected to have
estimated
$22.2\,\le\,\mu_{B,0}\,\le\,23.2$ 
mag\,arcsec$^{-2}$ and R$_{25}\,\ge\,16$ arcsec where $\mu_{\rm{B},0}$ denotes
the B band intrinsic disc central surface brightness, and  
R$_{25}$ denotes the major axis radius to
the 25 B mag\,arcsec$^{-2}$ isophote.
Our sample is by no means complete, but instead
is meant to explore the range of disc LSBG star formation histories.
Distances to the red LSBGs are unknown, however it is highly likely that they
are in the range 3000--8000 kms$^{-1}$ \cite{oneil1997}.  Our
comparison of the SFHs of red and blue LSBGs in section \ref{trans}
is strengthened if the red and blue LSBGs have similar physical sizes, however
even redshifts outside the above range will not significantly affect our
conclusions.

NIR $K'$ (1.94--2.29$\mu$m) passband images of these galaxies were
obtained in 1997 March 21 and 22, and 1998 May 5 using the Apache Point
Observatory 3.5-m
telescope.  Total on-source exposure times range from 16 minutes to 25
minutes.  Each observing sequence consisted of two sets of six 9.8 second
exposures on the object, bracketed on each side by a set of six 9.8
second sky exposures.
The data were dark subtracted and flat fielded using a median
combination of a given night's sky frames.  Both the dark subtraction
and flat fielding are accurate over large spatial scales to better than
$\sim\,0.2$ per cent of the sky level.  Sky subtraction was performed
using the weighted mean of the two nearest sky frames.  
The images were then registered using bright stars to determine the
offsets between frames, and the frames mosaiced.  Edges were visible
after mosaicing, thus a tip-tilt component was fit to the sky
background around the galaxy, and the fit subtracted.  
The addition of a tip-tilt component will not affect the
photometry of the galaxy in any individual frame.  The final images,
formed in this way, have residual background fluctuations of the order
of 0.03 per cent of the original sky level.  
Calibration was achieved using a number of
standard stars taken from Casali and Hawarden 
\shortcite{casali1992} at a similar range of airmasses to
our object frames, and is typically accurate to $\sim\,0.07$ mag.
Data for C3-2, taken on 1997 March 22, were taken during
non-photometric conditions and were calibrated to an accuracy of 0.05
mag during service time at the United Kingdom Infrared Telescope. 

Calibrated optical images in many of the required passbands were kindly
provided by Erwin de Blok for F561-1 and F568-3, Stacy McGaugh for
F563-V2 and Karen O'Neil for C3-2 and N10-2.  Additional $R$ band images of
C3-2 and N10-2 and $V$ and $R$ band images of F563-V2 were obtained on 1997
November 19 using the Isaac Newton 2.5-m Telescope as part of its
service programme.  Exposure times were 8 minutes in each passband.
The data were overscan corrected, bias subtracted and flat fielded
using sky flats.
The overscan correction
and bias subtraction were accurate to better than 0.1
per cent of the sky background.  Flat fielding accuracy is the limiting
factor in the depth of the optical surface photometry:
large-scale variations of $\sim\,1$ per cent of the sky background 
are observed and
these uncertainties are propagated into the error budget in the
subsequent analysis.  Calibration was achieved using one standard field
at a similar airmass from Landolt \shortcite{landolt1992}.  This
calibration is accurate to $\sim\,0.05$ mag in all passbands.

\section{Results}

\begin{figure*}
\begin{minipage}{175mm}
\begin{center}
  \leavevmode
  \epsffile{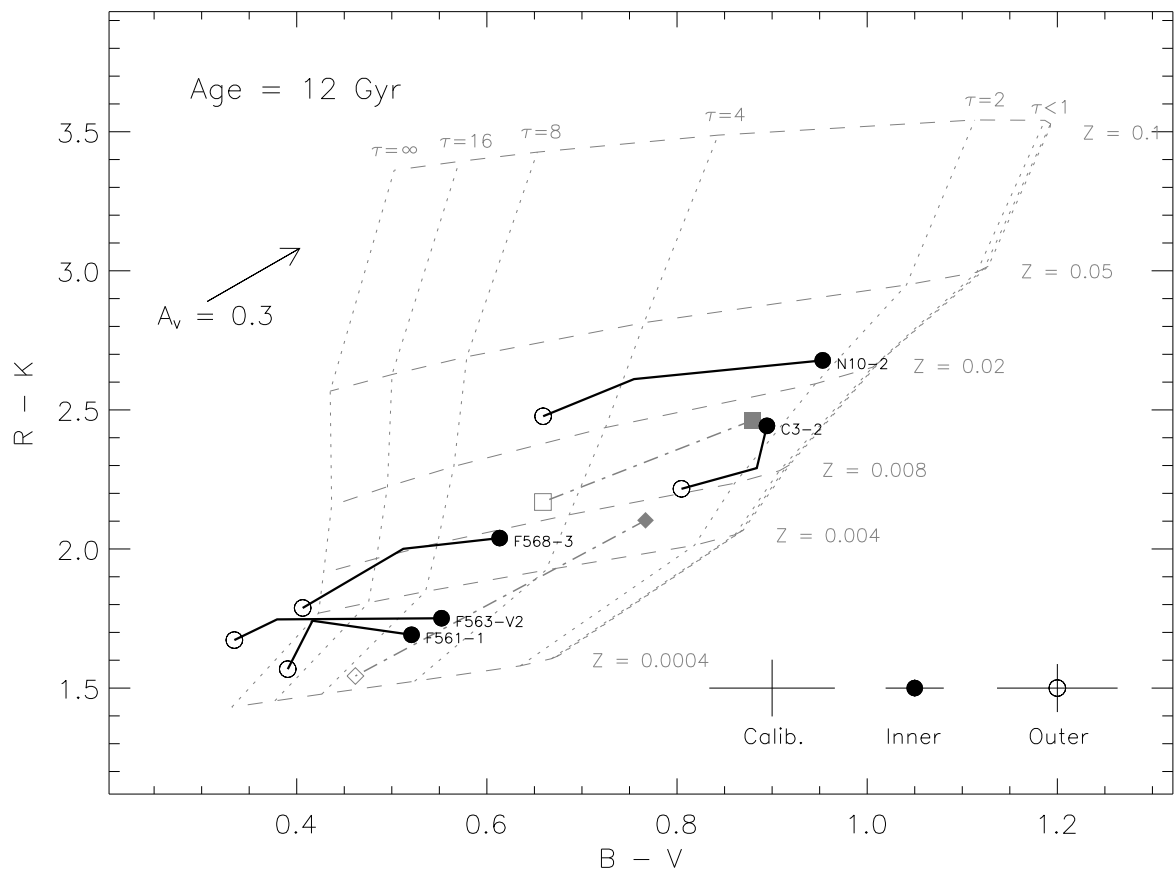}
\end{center}
\vspace{-0.3cm}
\caption{A colour-colour plot comparing the different stellar 
  populations in the red and blue LSBGs.
  The solid symbols denote central colours, and the open symbols the
  colours at 2 disc scale lengths.  The circles and solid lines
  are for the present sample of LSBGs.  Dot-dashed
  lines connect average colours of Sa--Sc (squares) and
  Sd--Sm (diamonds) galaxies from the sample of de Jong
  \protect\shortcite{dejong1996iv}.  Typical errors caused by sky level
  uncertainty are shown by the error bars in the lower
  right hand corner, where the central colour errors
  are denoted by the solid circle and those at
  2 disc scale lengths are denoted by the open circle.
  The calibration uncertainties are also shown.
  The model grid, based on the stellar population synthesis models of
  Bruzual and Charlot (in preparation), is discussed in the text. 
}
\label{fig:colcol}
\end{minipage}
\end{figure*}

Surface photometry was carried out using
the IRAF task {\it ellipse}.  
The centroid of the brightest region of the galaxy in $R$ band was used
as the galaxy centre.
The galaxy
ellipticity and position angle were determined from the $R$
band outermost isophotes.  
Estimation of the sky level used the outermost
regions of the surface photometry and the 
mean sky level in areas in the image that 
were free of extended galaxy emission
and contamination from starlight.  
Due to the
low surface brightness of our sample in all passbands, the error in the
sky level dominates the uncertainty in the 
photometry (this was also demonstrated using more realistic Monte Carlo
simulations, including the effects of seeing uncertainty, sky level
errors and shot noise).  This sky level, averaged over large
areas, is typically accurate
to a few parts in 10$^5$ for the $K'$ images, and better than $\sim\,0.5$ per
cent for the optical images.  
Galactic extinction corrections are from
Schlegel, Finkbeiner \& Davis \shortcite{sfd98}, and range between 0.07
and 0.20 mag in the $B$ band.
Galactic extinction corrected
galaxy colours in three radial bins ($0\,<\,r/h\,<\,0.5$, $0.5\,<\,r/h\,<\,1.5$
and $1.5\,<\,r/h\,<\,2.5$ where $h$ is the $R$ band disc scale length)
using images degraded to the same angular resolution are shown in
Fig. \ref{fig:colcol}.   

% Figure 1 lives here -- an optical-NIR colour-colour plot ---------------

It is clear that there are colour gradients in our
sample of LSBGs.  However, 
in order to interpret the colour gradients, it is necessary to compare
the data with the results of stellar population synthesis codes, such
as the GISSEL96 implementation of Bruzual \& Charlot (in preparation).  
In Fig. \ref{fig:colcol}, we use single metallicity stellar populations
with a Salpeter \shortcite{sp} IMF and 
an exponentially decreasing star formation rate characterised by an
e-folding timescale $\tau$.
  The dashed lines represent the colours of stellar populations with
  a fixed metallicity and a range of star-forming timescales from an
  instantaneous burst to a constant SFR.  The dotted
  lines represent the colours produced with a given star
  formation timescale and a range of metallicities.  The arrow denotes
  the dust reddening vector given by a screen model 
  using the extinction curve of Rieke \& Lebofsky 
  \shortcite{rieke1985} 
  for a visual extinction of 0.3 mag.
It should be noted that there is some uncertainty in the placement of
the model grid.  Charlot, Worthey \& Bressan \shortcite{charlot1996}
discussed the sources of error in stellar population synthesis models,
and concluded that the uncertainty in model calibration is about 0.06
mag in $B-V$ colour, and around 0.12 mag in $R-K$ colour, which is
comparable to the calibration error bar in Fig. \ref{fig:colcol}.

The observed colour gradients are consistent with a mean stellar
age gradient (parameterised by the exponential star forming 
timescale $\tau$), along with an expected contribution from 
metallicity effects.  The colour gradients may also have a
contribution from the effects of differential dust reddening:  this is
expected to be a small effect however and is discussed further in the
next section.
The existence of 
these colour gradients is insensitive to any zero point
uncertainties, and is very robust to reasonable flat 
fielding and sky level uncertainties.

From the difference in optical-NIR colours between the blue and red
LSBGs, the red LSBGs have considerably older mean
stellar ages than blue LSBGs, indicating an epoch of more vigorous star
formation.
Blue LSBGs, however, do not
necessarily need to be young (as we use a constant galaxy age of 12 Gyr
to create the model grid), but may retain their blue colours because of
a low roughly constant sporadic SFH \cite{deblok1997sfr}.
We address these stellar population differences further in section
\ref{trans}, where we ask if the two populations of LSBG 
have a common origin.
It should be noted that our conclusions are unchanged if other
combinations of optical-NIR colours are used. 

\section{Discussion}

\subsection{Are the stellar population differences real?}

We have interpreted Fig. \ref{fig:colcol} in terms of stellar
population differences, however we have so far neglected the effects of
differential dust reddening.
The difference between the red and blue LSBGs is 
consistent with our assumed form of dust reddening if $A_V\,\sim\,1$,
however this seems unlikely due to the large amount of smoothly
distributed internal reddening required, the reasonably smooth
morphologies of the red LSBGs, and the non-detection of either of the
red LSBGs in both the IRAS point source and small scale structure catalogues.
Also, the observed
colour gradients could be consistent with the effects of a gradient
in dust abundance in the LSBGs, with a central extinction of $A_V\,\sim\,1$
and little or no dust at 2 disc scale-lengths.  
This is by no means an implausible
amount of dust for a late type galaxy 
\cite{wk92,berlind1997}, however, there is evidence that LSBGs have less
dust  than HSB late type galaxies, namely the lack of strong dust
features in optical LSB galaxy images 
\cite{mcgaugh1994a,deblok1995}, the lack of detectable CO fluxes
\cite{schombert1990,deblok1997co}, 
and generally low ($A_B\,\sim\,1$ mag) Balmer decrements 
towards LSBG \hii regions \cite{mcgaugh1994b}.  Also,
the dust reddening vector shown in Fig. \ref{fig:colcol} was
made using the unphysical assumption of a foreground screen 
dust distribution.  More realistic modelling of the effects of dust
when mixed in with the stellar population results in `steeper' dust
reddening vectors \cite{dejong1996iv}.
However, these more realistic dust reddening 
vectors are relatively ill-constrained as the 
star/dust geometry, albedo (especially
NIR) of the dust particles and scattering phase function are all
uncertain.  Nevertheless, the colour gradients are likely to be
inconsistent with large amounts of dust reddening when more realistic
models are used.  

Uncertainties in the high mass end of the IMF do not alter these
conclusions.  
For example, use of a Scalo \shortcite{sc} or Miller \& Scalo 
\shortcite{ms} IMF changes only the high metallicity, large $\tau$ end
of the grid.  
While this could
change the absolute interpretation of colour gradients in terms of
values of metallicity and $\tau$, it is still possible to spot relative
metallicity and age trends in and between galaxies, and remain
largely unaffected by this uncertainty.  
Note, however, that {\it large} changes in the IMF could significantly affect
our relative age estimates.
For example, explanation of the optical-NIR colours of the red LSBGs in
terms of an IMF heavily biased towards the formation of low mass stars
is possible, as the optical-NIR colours in essence indicate that
red LSBGs simply lack high mass stars.  In this case, the age of the
red LSBGs would be much more difficult to constrain.

\subsection{Do red and blue LSBGs share common ancestors?} \label{trans}

\begin{figure*}
\begin{minipage}{175mm}
\begin{center}
  \leavevmode
  \epsffile{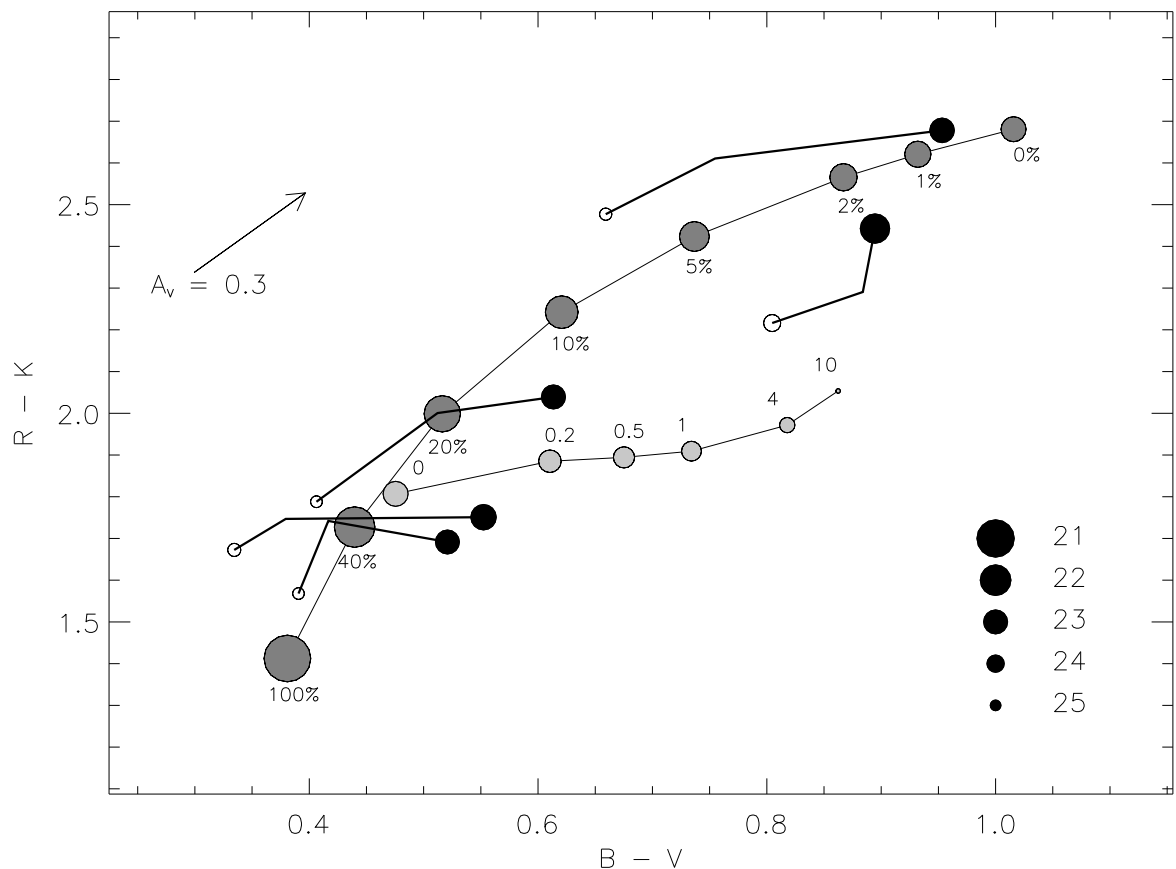}
\end{center}
\vspace{-0.3cm}
\caption{A plot showing the possibilities for transformation between
  red and blue LSBGs.  
  Central and outer colours for our sample are denoted by solid and
  empty dark circles, with their sizes indicating the $B$ band surface
  brightnesses at those radii.  The upper track (dark gray) 
  shows the colours of a
  12 Gyr old solar metallicity stellar population.  Added to it is a
  1.25 Gyr old single burst population with $Z\,=\,0.004$ with mass
  fractions ranging from 0 to 100 per cent.  The lower track (light
  gray) shows
  colours of a $\tau\,=\,16$ Gyr model with an age of 12 Gyr, with the
  star formation truncated between 0 and 10 Gyr ago.  Plot symbol size
  denotes changes in $B$ band surface brightness in mag\,arcsec$^{-2}$.
}
\label{fig:evol}
\end{minipage}
\end{figure*}

Fig. \ref{fig:colcol} suggests that the red LSBGs have
undergone a period of more vigorous star formation (both from the
dominance of the older stellar populations, and from the inferred 
stellar metallicities), whereas blue LSBGs are well described by
stellar populations with ongoing $\la\,1/3$ solar metallicity 
star formation.
However, is it possible that the red and blue LSBGs share the same
origin:  that is, can red and blue LSBGs transform from one into the
other readily?  
In Fig. \ref{fig:evol}, we address this question.

Our red LSBGs, using the models of Bruzual \& Charlot (in preparation), 
have the optical-NIR colours of an old stellar 
population with roughly solar 
metallicity.  If one adds 20 to 30 per cent, by mass, of
a young $\sim\,1$ Gyr stellar population with low metallicity $Z\,\la\,
0.004$ it is possible to reproduce the colours of the blue LSBGs.
However, the addition of young stars at all radii (to reproduce the
colour of blue LSBGs at all radii) will increase the blue surface
brightness by $\sim\,-2$ mag, giving central surface brightnesses of 
$\sim\,21$ mag\,arcsec$^{-2}$ for these `transformed' red LSBGs.  Thus,
it is impossible to transform a red LSBG into a blue LSBG due to
surface brightness constraints.

Alternatively, to transform a blue LSBG into a red one, the SFH must be
truncated.
Truncation of star formation 
will cause the optical colours to redden (note however that
the stellar metallicity would still appear lower than those observed in
our red LSBG: see Fig. \ref{fig:evol}) at the expense of
surface brightness (dimming of $\sim\,1.5$ mag in $B$, and $\sim\,0.8$
mag in $K$ is expected after 4 Gyr, when compared to a similar galaxy at
the time of truncation which is still forming stars with $\tau\,\sim\,
16$ Gyr).  
Thus, it is clear that the blue LSBGs cannot reproduce {\it our} sample
of red LSBGs not only because of metallicity constraints, but also
because the surface brightness would be too faint.  Fading from a roughly
solar metallicity HSB galaxy to a red LSBG may be possible: 
measuring the \hi gas fractions of red
LSBGs would test this idea, 
as significant fading is unlikely in gas-rich galaxies.

The stellar metallicities implied by the optical-NIR colours are
consistent with those inferred from the optical colours alone, 
galaxy type and \hii region metallicity (where available).
Comparison of the blue LSBGs with the
sample of de Jong \shortcite{dejong1996iv} indicates stellar
metallicities similar to many Sdm galaxies.  The implied stellar
metallicities are comparable to those seen in blue LSBG \hii
regions \cite{mcgaugh1994b,ronnback1995,deblok1997ab}.  
In addition, the red LSBGs have roughly solar
inferred stellar metallicities, which are compatible with
the Sa galaxies in de Jong \shortcite{dejong1996iv}.

Padoan, Jiminez \& Antonuccio-Delogu \shortcite{padoan1997}
proposed that the blue optical
colours of blue LSBGs were due to very low stellar metallicity ($Z\,\sim\,
0.0002$), old single burst stellar populations.
By inspection of Fig. \ref{fig:colcol}, it is clear that the 
optical-NIR colours of blue LSBGs imply
stellar metallicities typically {\it a factor of 20} larger than those
required to fit their optical colours with an old, single burst 
stellar population.  
This higher metallicity necessarily implies more
recent star formation to make the stellar colours bluer, therefore
necessitating younger mean stellar ages.  
However, it should be stressed that blue LSBGs need not be young, but they
must have substantial recent star formation.
As stated in 
Padoan et al.\ \shortcite{padoan1997} and borne out by our own tests,
this conclusion is robust to changes in the stellar IMF and the details
of the overall star formation history.

\section{Conclusions}

As part of an ongoing study, we have obtained $K'$ images of five red
and blue LSBGs.
With the addition of optical $B$, $V$
and $R$ images, we found the following.
\begin{itemize}
\item Optical-NIR radial colour gradients are present 
  in red and blue LSBGs, consistent with mean stellar age gradients,
  with the outer regions of LSBGs being younger than the central regions.
\item Very low stellar metallicities are ruled out for these galaxies,
  which is inconsistent with the SFH proposed by
  Padoan et al.\ \shortcite{padoan1997}.
\item Red LSBGs have the optical-NIR colours of passively-evolving 
  roughly solar
  metallicity stellar populations.  In contrast, blue LSBGs are still
  actively forming stars, albeit at a low overall rate.
\item These results suggest that red and blue LSBGs are two
  different types of galaxy, and 
  represent two independent routes to low $B$ band surface
  brightness:  the blue LSBGs are well described by models with a low,
  roughly constant SFRs, whereas the red LSBGs are more consistent with
  a `faded HSB disc' scenario.
\end{itemize}

\section*{Acknowledgements}

We would like to thank Erwin de Blok, Karen O'Neil and 
Stacy McGaugh for providing
surface photometry and images of galaxies in their sample, and for
helpful discussions.  In particular, we would like to thank Karen
O'Neil for providing the coordinates of her LSBG sample before their
publication. 
We would also like to thank the referee for useful comments on the manuscript.
EFB would like to thank the Isle of Man Education Department for their
generous support.
Support for RSdJ was provided by NASA through Hubble Fellowship
grant \#HF-01106.01-98A from the Space Telescope Science Institute,
which is operated by the Association of Universities for Research in
Astronomy, Inc., under NASA contract NAS5-26555.
We also thank the
U. Chicago TAC for regular time allocations to make the
$K'$ observations.  
Some of the observations described in this letter were made during
service time at the Isaac Newton Telescope and at the United Kingdom
Infrared Telescope.  This project made use of STARLINK facilities in Durham.

\end{document}